\documentclass[10pt,twocolumn,letterpaper]{article}

\usepackage{cvpr}
\usepackage{times}
\usepackage{epsfig}
\usepackage{graphicx}
\usepackage{amsmath}
\usepackage{amssymb}
\usepackage{color}
\usepackage{booktabs,threeparttable}
\usepackage{colortbl}
\usepackage[export]{adjustbox}
\usepackage{caption}
\usepackage{subfigure}
\usepackage{textcomp}

\cvprfinalcopy 

\def\cvprPaperID{} 

\begin{document}

\title{Automatic Gain Control Design for Dynamic Visible Light Communication Systems}

\author{Yingwen Zhang\\
University of Science and\\
Technology of China, Hefei\\
\and
Xianqing Jin\\
University of Science and\\
Technology of China, Hefei\\
\and
Weibin Jiang\\
University of Science and\\
Technology of China, Hefei\\
\and
Xinmin Chen\\
University of Science and\\
Technology of China, Hefei\\
\and
Zhengyuan Xu\\
University of Science and\\
Technology of China, Hefei\\
}

\maketitle

\begin{abstract}
  For dynamic visible light communication (VLC) systems, received optical power fluctuates largely due to the fast movement of VLC terminals. Such fluctuations will cause possible signal clipping and quantization noise at the analog-to-digital converters (ADCs) and thus aggravate signal-to-noise ratio (SNR) and reliability of the communication link. To mitigate this effect, the automatic gain control (AGC) technique is often introduced in the receiver front-end to adaptively adjust the electrical signal strength. 
  
  In this paper, we provided a VLC front-end analysis considering AGC model. Based on the model, some design principles of AGC are theoretically derivated: the effects of AGC index $m$, AGC maximum gain $g_{max}$ and AGC equilibrium range $DR$. Next, an analog AGC amplifier was carefully implemented and the effectiveness of our AGC modeling was proved through BER experiments. Finally, real-time 25Mb/s on-off-keying (OOK) experiments with AGC function were demonstrated in a dynamic link with different speeds. Experimental results show that the AGC can stabilize the BER performance and improve the system performance in the speed of 1m/s.

\end{abstract}


\section{Introduction}
As an emerging technology, visible light communication (VLC) use the license-free light spectrum to provide high speed communication and network connectivity, which can appreciably meet the unprecedented demands of communication and connection in the era of Internet of things \cite{wu2014visible}. 

Over the decades, the user mobility problem is one of the challenges that both the research community and industry have to tackle to make VLC really commercial. On the account of high directionality and attenuation of light, in mobile scenarios, the received signal amplitude is easily either too large causing ADC clipping or too small resulting in quantization noise, and thus BER deterioration. Similar to a mature RF system, the solution to this is to introduce automatic gain control (AGC), which can stabilize the input amplitude of ADC regardless of the signal variation, into the VLC system\cite{cuailean2017current}.

For AGC, the signal strength is detected and then used to control the signal gain, where the strength detection process can be accomplished either digitally, namely, ADC sampling and digital calculation, or with analog circuit. Digital AGCs are widely used in RF systems for its  flexibility of adjustment and high reliability, where the ADC quantizer combined with AGC were analyzed in \cite{yeh2004agc,sun2011particle,narieda2013agc}. While in the VLC research, \cite{okada2009vehicle} first realized the importance of AGC while studying the VLC vehicular communication. In \cite{cailean2013robust}, the authors conducted preliminary experiments of AGC for VLC of different distances, where the signal gain was set by switches. Due to the ability to mitigate the amplitude attenuation, an analog AGC amplifier was used to extend the LED bandwidth in \cite{chow2015practical}. Besides, in terms of AGC circuit design towards VLC system, certain papers were found\cite{fuada2017automatic,zhang2008analog}.

Until now, effects of AGC has not been fully studied in VLC situations, both theoretically and experimentally. In this paper, starting from a simple amplifier model, we derived the SNR model for AGC. Besides, combined with VLC front-end reception model, some AGC practical design insights for VLC systems were given. Finally, experiments were conducted to evaluate the effectiveness of AGC.  
\begin{figure}[!htbp]
	\begin{center}
		\includegraphics[height=3.1cm]{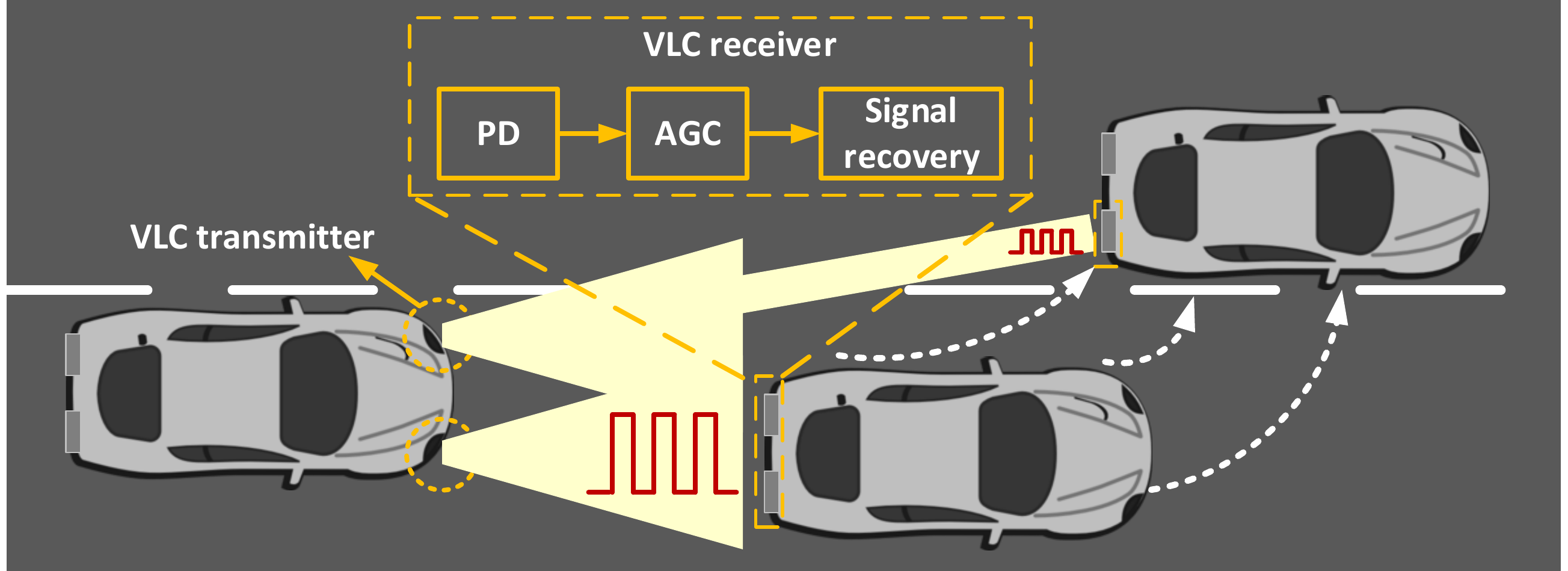}
		\makeatletter\def\@captype{figure}\makeatother
		\caption{A typical mobile VLC scenario}\label{fig:sysBlocks}
	\end{center}
\end{figure}
\vspace{-0.5cm}
\section{System Model}
\subsection{Signal propagation} \label{suc:A}
For the VLC transmitting-end, signals are firstly generated by the programmable hardware, e.g., FPGA. Then proper power amplification and DC bias are given to drive the LED. In practice, due to the DAC quantization noise and the power amplifier noise, we consider that the transmitted signals are noisy, and for a certain discrete time, the transmitted optical signal of LED was denoted by random variable
\begin{equation}\label{eq:LED_output}
b=\alpha(s+n_t+v_b)
\end{equation}
where $b$ and $s$ are random variables, representing the optical power and electrical signal of transmitter; $n_t\sim N(0,\sigma^2_t)$ is the transmitter noise, assumed as zero-mean Gaussian white noise and to be independent of $s$; $v_b$ is the DC bias voltage; $\alpha$ is the electro-optical conversion coefficient. 

Through channel propagation, the converted electrical current at the receiving-end is\cite{Kahn1994Wireless}
\begin{equation}\label{eq:APD_output}
x=\beta(hb+b_0)+n_i+n_d
\end{equation}
where $x$ is the electrical current of front-end; $\beta$ is the APD responsivity; $b_0$ is the optical power of ambient lights, which usually is considered as a constant; $h$ is the Lambertian channel gain, given by \cite{Kahn1994Wireless}
\begin{equation}\label{eq:h}
h=
\begin{cases}
\displaystyle\frac{(n+1)A}{2\pi d^2}cos^n\phi T_s(\psi)g(\psi)cos\psi & 0\leqslant\psi\leqslant\Psi_c \\
\displaystyle 0 & \Psi_c\leqslant\psi
\end{cases}
\end{equation}
where the Lambertian index $n$ depends on the half-power angle of transmitter $\phi_{1/2}$, given by $n=-1/log_2(cos\phi_{1/2})$; $A$ is the receiver area of photodiode; $d$ is the communication distance; $\phi$ is the emission angle; $\psi$ is the incident angle; $T_s(\psi)$ is the optical filter gain; $\Psi_c$ is the angle of half field of view (FOV);  $g(\psi)=\chi^2/sin^2\Psi_c$ is the concentrator gain, where $\chi$ is the refractive index. 

In (\ref{eq:APD_output}$), n_i \sim N(0,\sigma^2_i)$ is the input-independent noise, representing both the circuit noise and the ambient light induced shot noise, given by 
\begin{equation}\label{eq:idpnoise}
\sigma^2_i=\sigma^2_{c}+2qMF_A\Delta f\beta b_0
\end{equation}
where $\sigma^2_{c}$ is the circuit noise; $q$ is the electron charge; $M$ is the APD multiplication factor; $F_A$ is the excess noise factor; $\Delta f$ is the system bandwidth.
$n_d \sim N(0,\sigma^2_d)$ is the input-dependent noise, representing the signal induced shot noise, whose average variance $\bar{\sigma^2_d}$ can be denoted by\cite{Ming2015Coding}
\begin{equation}\label{eq:shotNoise}
\bar{\sigma^2_d}=\mathbb{E}_s[\sigma^2_d]=
2qMF_A\Delta fh\alpha\beta(\mathbb{E}[s]+v_b)
\end{equation}
Considering DC components are blocked by capacitors, the total front-end electrical AC power is thus given by
\begin{equation}\label{eq:frontEndElecPower}
\begin{split}
&p_x=\mathbb D[x]r_l \\
&=\{\mathbb D[\beta hb]+\bar{\sigma^2_d}+2\beta h\operatorname{Cov}[b,n_d]+\sigma^2_i\}r_l
\end{split}
\end{equation}
where $p_x$ represents the converted electrical power; $r_l$ is the front-end load resistor. Despite the signal dependency between $b$ and $n_d$, the covariance term $\operatorname{Cov}[\cdot]$ is $0$. Using (\ref{eq:LED_output}), (\ref{eq:APD_output}) and (\ref{eq:shotNoise}), we have
\begin{equation}\label{eq:finalPowerEq}
p_x=p_s+p_n \\
\end{equation}
where $p_s$ represents the electrical power of signal and $p_n$ is the total electrical power of front-end noise, denoted by
\begin{gather}\label{eq:ps}
p_s=(h\alpha\beta)^2\mathbb{D}[s]r_l \\
p_n=\lambda p_s+2qMF_A\Delta f\sqrt{(p_sr_l)/\mathbb{D}[s]}(\mathbb{E}[s]+v_b)+\sigma^2_ir_l \label{eq:pn}
\end{gather}
where $\lambda$, defined as $\sigma^2_t/\mathbb{D}[s]$, is the transmitted noise signal ratio. The transmitter is noiseless once $\lambda=0$. 

The front-end input SNR thus is denoted by
\begin{equation}\label{eq:SNR}
SNR_i=\frac{p_s}{p_n}
\end{equation}
\subsection{Automatic gain control}
\begin{figure}[!htbp]
	\vspace{-0.4cm}
	\begin{center}
		\includegraphics[height=4cm]{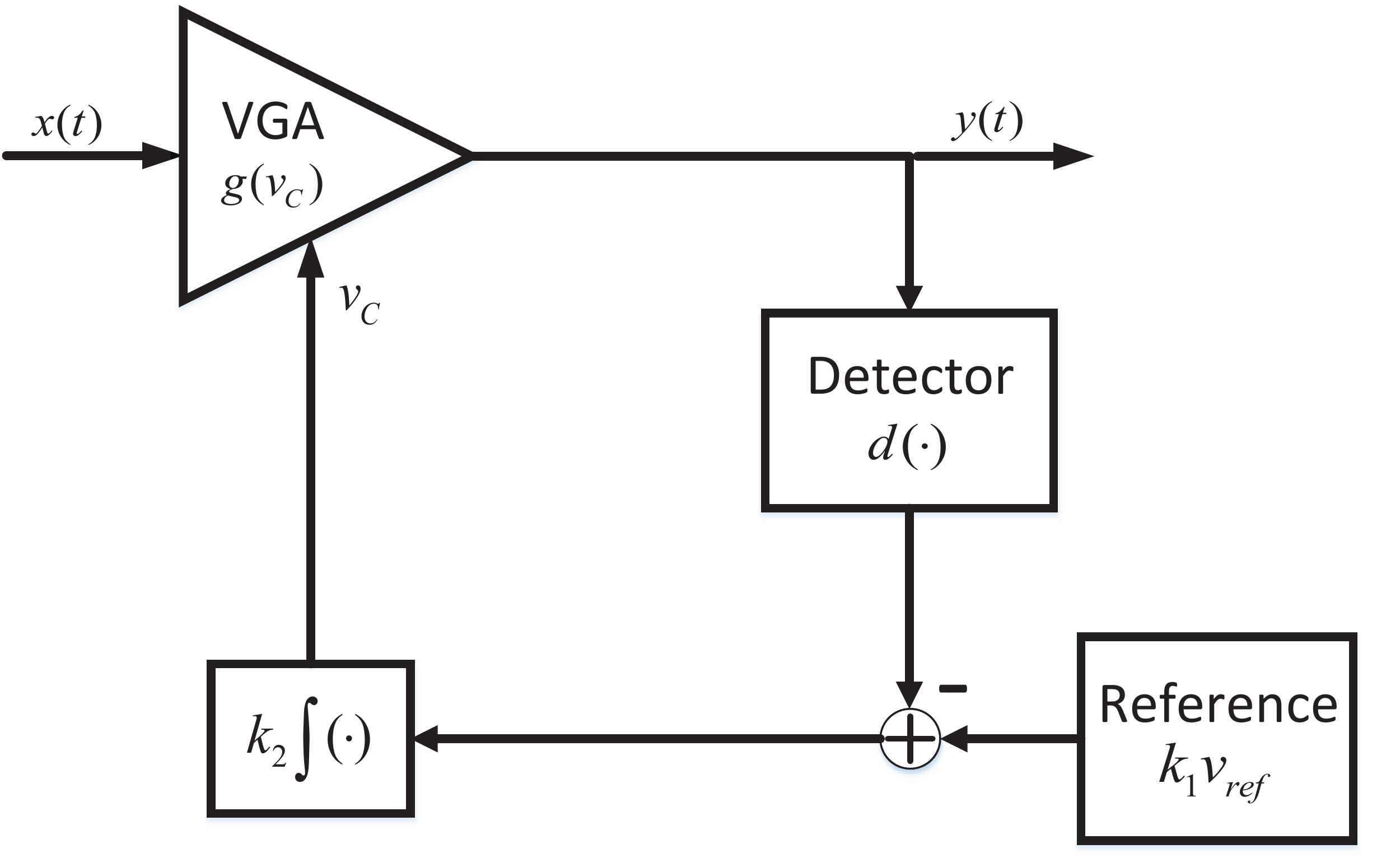}
		\makeatletter\def\@captype{figure}\makeatother
		\caption{A typical feedback AGC diagram}\label{fig:AGC}
	\end{center}
\end{figure}

After the front-end reception, signals are fed to AGC to prevent signal fluctuations caused by relative movements of the VLC transmitter and receiver. The essence of an AGC amplifier is a variable gain amplifier (VGA) that can adjust gains dynamically according to the input strength, signal amplitude or power, so that the output remains stable (see AGC diagram in Fig.(\ref{fig:AGC})). Here $x(t)$ and $y(t)$ are the input and output signal of continuous time, respectively. $g(v_c)$ is the VGA gain, and usually is a exponential function of control voltage $v_c$, leading a constant setting time of AGC loop that is independent of input signal $x(t)$\cite{Prez2011Automatic}. The VGA output strength is then detected by a loop detector, whose output $d(y(t))$ is further compared with a external reference voltage $v_{ref}$. Finally, through a loop filter, the comparison difference is smoothed into a DC voltage $v_c$ to control the VGA's gain, where we have \cite{Prez2011Automatic}
\begin{equation}\label{eq:loopDynamic}
v_c(t)=k_2\int_{0}^{t}(k_1v_{ref}-d(y(\tau)))\, d\tau
\end{equation}  
where $k_1$ and $k_2$ are the circuit scaling factors. The strength detection can be implemented digitally using ADC or by analog detectors, such as the square-law detector\cite{whitlow2003design}. 

Apparently, from (\ref{eq:loopDynamic}), as the input strength rises, the detector output becomes larger than the reference level, namely, $k_1v_{ref}-d(y(\tau))<0$. Through the time integral, then $v_c$ will decline correspondingly, resulting in a smaller amplifier gain $g(v_c)$, which will lower the output $y(\tau)$. Subsequently, the loop reaches the equilibrium-state as $k_1v_{ref}=d(y(\tau))$ and outputs a constant power that can be adjusted by $v_{ref}$. This process of loop settling is the same vice versa when the input strength declines.

Though such a AGC system is inherently non-linear, for small changes of input strength, after logarithmized formulation with proper approximation technique, namely, Taylor series expansion, we can operate and analyze the AGC system as a first order linear system in decibel (dB)\cite{khoury1998design}. 

The dynamic performance of first-order linear system is described by the time constant $\tau$\cite{kuo1995automatic}. For instance, for a unit step input, it takes $t=3\tau$ for the system to rise from 0\% to 95\% of the final steady amplitude. In practical mobile VLC scenarios, the designed time constant thus should be much smaller than the power fluctuation period.  

Next, we give a signal SNR model for the settled AGC amplifier. Basically, for a fixed gain amplifier, after a power amplification of $g$, we have\cite{niknejad2007electromagnetics}
\begin{equation}\label{eq:ampModel}
p_y=gp_x+p_a  
\end{equation}
where $p_a$ is the AGC noise power, which is independent of the input. Using (\ref{eq:finalPowerEq}), we have the output SNR
\begin{equation}\label{eq:SNRo}
SNR_o=\frac{gp_s}{gp_n+p_{a}}=\frac{SNR_i}{1+p_{a}/gp_n}
\end{equation}
As discussed earlier, the output of the equilibrium-state AGC is a stable power $p_e$. While in practice, on account of the limit gain range of the VGA component, once the gain that needed to amplify the input to the equilibrium level is out of range, the AGC amplifier will behave as a fixed gain amplifier with the boundary gain $g_{max}$ or $g_{min}$. Thus $g$ was denoted by a piecewise function
\begin{equation}
g=
\begin{cases}\label{eq:AGCgain}
g_{max} & p_{x}<p_{l} \\
\displaystyle\frac{p_e-p_{a}}{p_{x}} & p_{l}\leqslant p_{x}\leqslant p_{u} \\
g_{min} & p_{x}>p_{u}
\end{cases}
\end{equation}
where $g_{max}$ and $g_{min}$ are the maximum and minimum accessible gain of VGA; $p_e$ is the equilibrium output power of AGC; two equilibrium thresholds $p_{l}$ and $p_{u}$ were given by
\begin{gather}\label{eq:Pth1}
p_{l}=\frac{p_e-p_{a}}{g_{max}} \\
p_{u}=\frac{p_e-p_{a}}{g_{min}} \label{eq:Pth2}
\end{gather}

Applying (\ref{eq:AGCgain}) to (\ref{eq:SNRo}), defining the AGC index $m=p_e/{p_{a}}$, we have the output SNR of AGC
\begin{equation}\label{eq:AGCTotalSNRo}
SNR_o=
\begin{cases}
\displaystyle\frac{SNR_i}{1+p_a/g_{max}p_n} & p_x< p_l \\
\displaystyle\frac{(m-1)SNR_i}{m+SNR_i} & p_l\leqslant p_x\leqslant p_u \\
\displaystyle\frac{SNR_i}{1+p_a/g_{min}p_n} & p_x< p_u
\end{cases}
\end{equation}

\section{Analytical Results and Analysis}
\subsection{Effect of AGC index $m$}\label{subsec:1}
\begin{figure}[!htbp]
	\begin{center}
		\includegraphics[height=6cm]{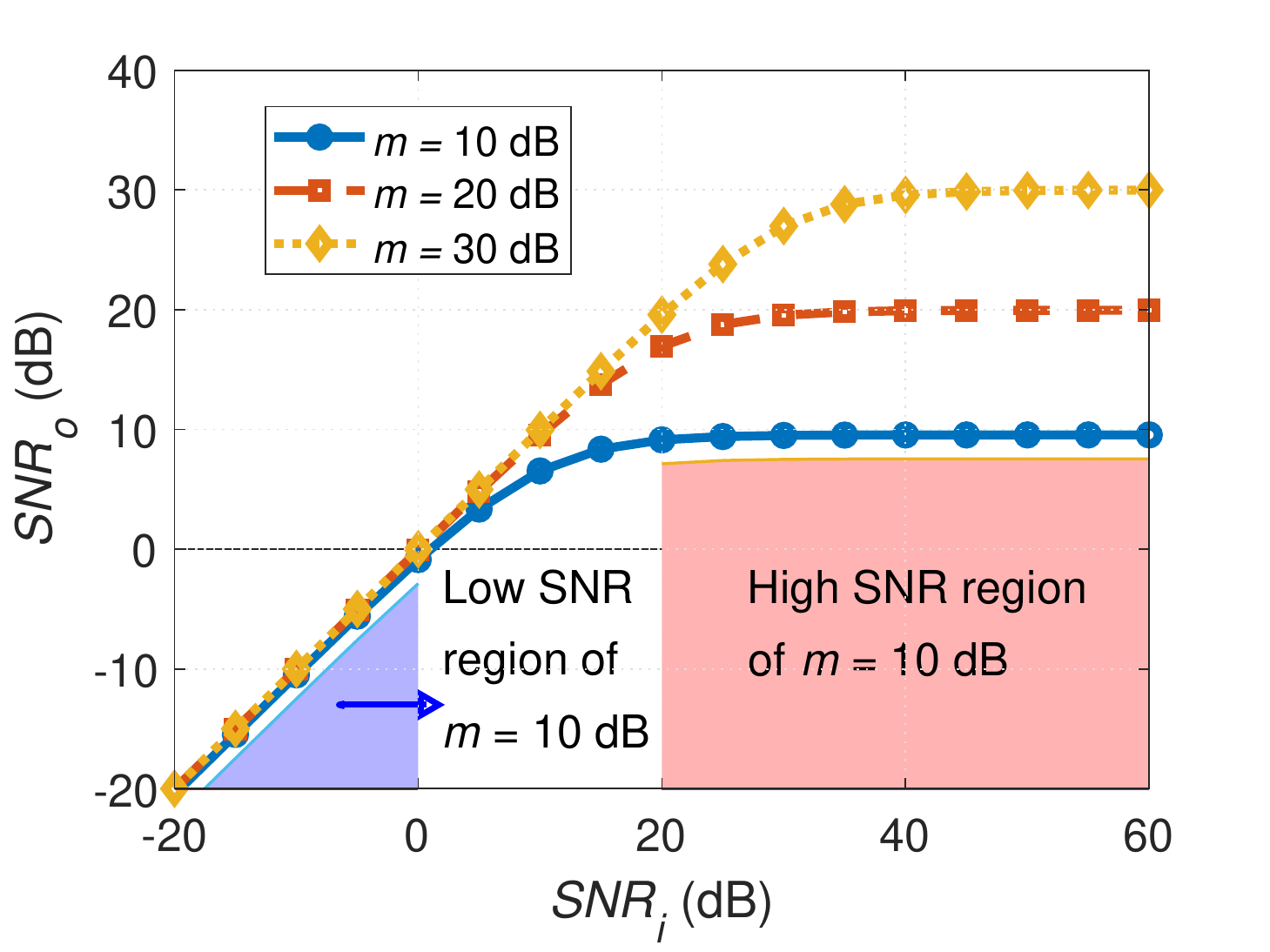}
		\makeatletter\def\@captype{figure}\makeatother
		\caption{AGC SNR performance with respect to different $m$}\label{fig:AGC_ioSNR}
	\end{center}
\end{figure}

As shown in (\ref{eq:AGCTotalSNRo}), when AGC is in equilibrium-state ($p_l\leqslant p_x\leqslant p_u$), the output SNR is determined by both the input SNR and the AGC index $m$.  In Fig.\ref{fig:AGC_ioSNR}, we first investigated the output SNR performance of equilibrium-state AGC with different $m$, where $m$ was presented in dB like SNR.

It should be noted in the beginning that, $SNR_o$ is monotonically increasing with respect to $m$. As illustrated in Fig.\ref{fig:AGC_ioSNR}, the SNR curve can be analyzed from three different regions (see the example of region under curve of $m$ = 10 dB). Firstly, in the low SNR region, namely, $SNR_i\ll m$, the output SNR approaches to the input SNR, and is barely improved as we increase $m$. On the contrary, in the high SNR region, namely, $SNR_i\gg m$, certain SNR floors of $m - 1$ occur.

Overall, while the AGC is in equilibrium-state, the system SNR benefits from a larger $m$, avoiding the floor effect. From definition, large $m$ means increasing $p_e$. Nevertheless, the AGC equilibrium thresholds, i.e., $[p_{l},p_{u}]$, is controlled by $p_e$ (see (\ref{eq:Pth1}) and (\ref{eq:Pth2})), simply increasing $p_e$ will result in higher thresholds, which implies a possible trade-off while designing practical systems.

\subsection{Effect of VGA gain $g_{max}$}
\begin{figure}[!htbp]
	\begin{center}
		\includegraphics[height=6cm]{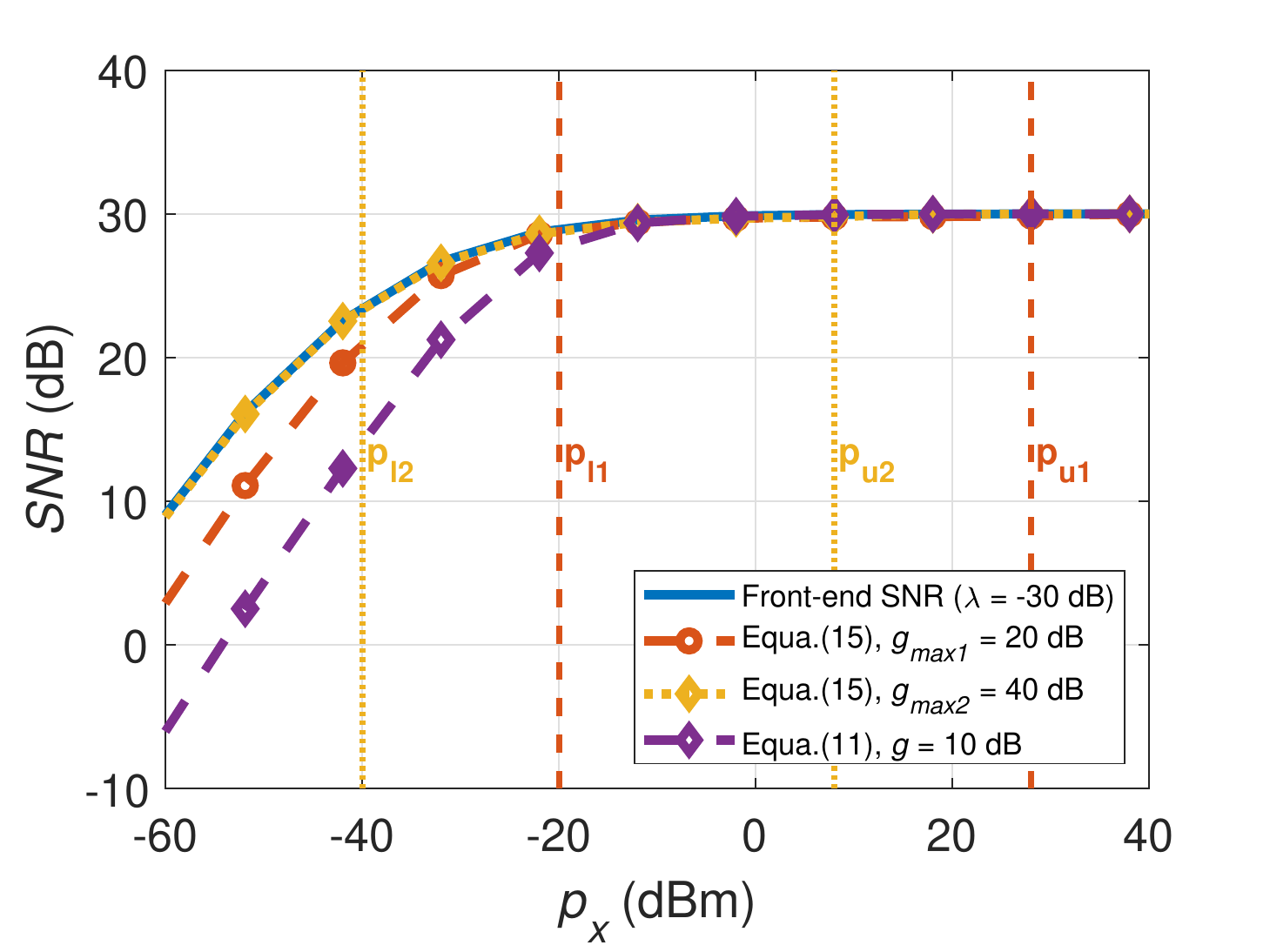}
		\makeatletter\def\@captype{figure}\makeatother
		\caption{AGC output SNR performance with different VGA maximum gain}\label{fig:CombinedAGC_gmax}
	\end{center}
\end{figure}

\begin{table}
	\centering
	\caption{Simulation Parameters}\label{tb:Simulation Parameters}
	\begin{tabular}{|c|c|c|c|}
		\hline        
		Transmitted signal power ($\mathbb{D}[s]$)&0.08 A$^2$  \\ \hline
		LED DC bias ($v_b$)&8 V \\ \hline
		LED threshold voltage&6 V \\ \hline
		Transmitted noise signal ratio ($\lambda$)&-30 dB \\ \hline
		LED conversion coefficient ($\alpha$)&0.125 A/W \\ \hline
		Electron charge ($q$)&$1.6\times10^{-19}$ C \\ \hline
		Multiplication factor ($M$)&30 \\ \hline
		Excess noise factor ($F_A$)&4.77 \\ \hline
		System bandwidth ($\Delta f$)&12.5 MHz \\ \hline
		Independent noise density&$6.654\times10^{-15}$ mW/Hz \\ \hline
		AGC noise density&$2.71\times10^{-12}$ mW/Hz \\ \hline
		Load resistor ($r_l$)&$50 \Omega$ \\ \hline
		VGA adjustable gain range&48 dB \\ \hline
		AGC equilibrium output ($p_e$)&0 dB \\ \hline
		APD responsivity  ($\beta$)& 460 W/A \\ \hline
		Half-power angle ($\phi_{1/2}$)& 60 degree \\ \hline 
		Optical filter gain ($T_s(\psi)$)& 1 \\ \hline
		Half-FOV ($\Psi_c$)& 60 degree \\ \hline
		Reflective index ($\chi$)& 1.5 \\ \hline
	\end{tabular}
\end{table}
Next, performances of the AGC output SNR with different VGA maximum gains $g_{max}$ under certain front-end situation was given (see Fig.\ref{fig:CombinedAGC_gmax}). In this simulation, the transmitted signal is the zero-mean bipolar OOK, where $\mathbb E[s]=0$. The transmitted noise signal $\lambda$ was set to -30 dB without loss generality. APD parameters are with Hamamatsu C12702-04 and we measured the module open-circuit output in indoor with curtains open to approximate the noise power term $\sigma^2_ir_l$. Also, the AGC amplifier open-circuit output was measured as the AGC noise $p_{a}$. The VGA gain range is fixed by 48 dB, where the VGA minimum gain is then $g_{min}=g_{max}-48$ in dB. The $p_e$ was set to 0 dBm considering the actual ADC input limits (detail system parameters see Table \ref{tb:Simulation Parameters}).

By simple calculation, we have $m$ = 44.7 dB and thresholds were plotted with vertical lines. In Fig.\ref{fig:CombinedAGC_gmax}, the input SNR of AGC is the front-end SNR calculated by (\ref{eq:finalPowerEq})-(\ref{eq:SNR}). Here, the front-end SNR is upper bounded by the transmitted SNR $1/\lambda$ as we included transmitter noise.

When $p_{x}<p_1$, the output SNR is improved by increasing $g_{max}$. In this interval, AGC behaves as a fixed gain amplifier with gain $g_{max}$. The output SNR described in (\ref{eq:SNRo}), is mainly determined by the term $p_a/gp_n$ while given the input SNR. In fact, since the front-end input noise power $p_n$ is lower bounded by the term $\sigma_c^2r_l$, careful $g_{max}$ design 
\begin{equation}\label{eq:gmaxDesign}
g_{max}\gg\frac{p_{a}}{\sigma_c^2r_l}=26.1\;dB
\end{equation}
make $p_{a}/(g_{max}\sigma_i^2r_l)\rightarrow0$ and thus $SNR_i\approx SNR_o$ (see the curve of $g_{max2} = 40$ dB in Fig.\ref{fig:CombinedAGC_gmax}). In addition, aside from the SNR loss caused by small gain, it is noteworthy that as $p_x$ decreases, quantization noise will occur at ADCs stage. 

When $p_1\geq p_{x}\geq p_2$, since the input SNR is bounded by 30dB, where $SNR_i\ll m$, the low SNR region conclusion can be applied in this interval. When $p_{x}>p_1$, AGC is a fixed gain amplifier with the bounded input SNR. Same analysis can be applied to this situation using (\ref{eq:SNRo}). Besides, a amplifier with fixed gain of 10 dB was plotted as reference in Fig.\ref{fig:CombinedAGC_gmax}.

\subsection{Analysis of channel dynamic range}
Furthermore, the dynamic range of the received optical power corresponding to the AGC equilibrium range were analyzed. From Section.\ref{suc:A}, ignoring the ambient light  $b_0$, the average received optical power is denoted by
\begin{equation}
\bar{o}=\mathbb{E}[hb]=\alpha hv_b
\end{equation}

In our simulation, considering the SNR of the equilibrium thresholds $p_l$ and $p_u$ are higher than 10dB (see Fig.\ref{fig:CombinedAGC_gmax}), we have $p_x\approx p_s$. Using (\ref{eq:ps}), the AGC equilibrium range can thus be noted by
\begin{equation}
DR=10\,lg\frac{p_u}{p_l}\approx 20\,lg\frac{h_u}{h_l}
\end{equation}
where $DR$ is the equilibrium range of AGC; $h_u$ and $h_l$ are the channel gain when the received electrical power are $p_u$ and $p_l$, respectively. The dynamic range for the received optical power is then calculated by
\begin{equation}
10\,lg\frac{\bar{o_u}}{\bar{o_l}}=10\,lg\frac{h_u}{h_l}=\frac{1}{2}DR=24\;dB
\end{equation}
To be more specific, using (\ref{eq:h}), (\ref{eq:ps}) and parameters given in Table.\ref{tb:Simulation Parameters}, considering the emission and incident angle are both 90 degree, the distance $d_l$ corresponding to $p_l$ is calculated as 1.57 m. And considering distance at 1 m, the deviation angle $\phi_l$ corresponding to $p_l$ is calculated as 66 degree, covering the whole FOV of 60 degree.  

\section{Experimental Results}
\subsection{AGC experimental transmission characteristic}
\begin{figure}\centering
	\subfigure[System setup for AGC transmission and BER experiments]{
		\begin{minipage}[b]{0.9\columnwidth}
			\includegraphics[height=3cm]{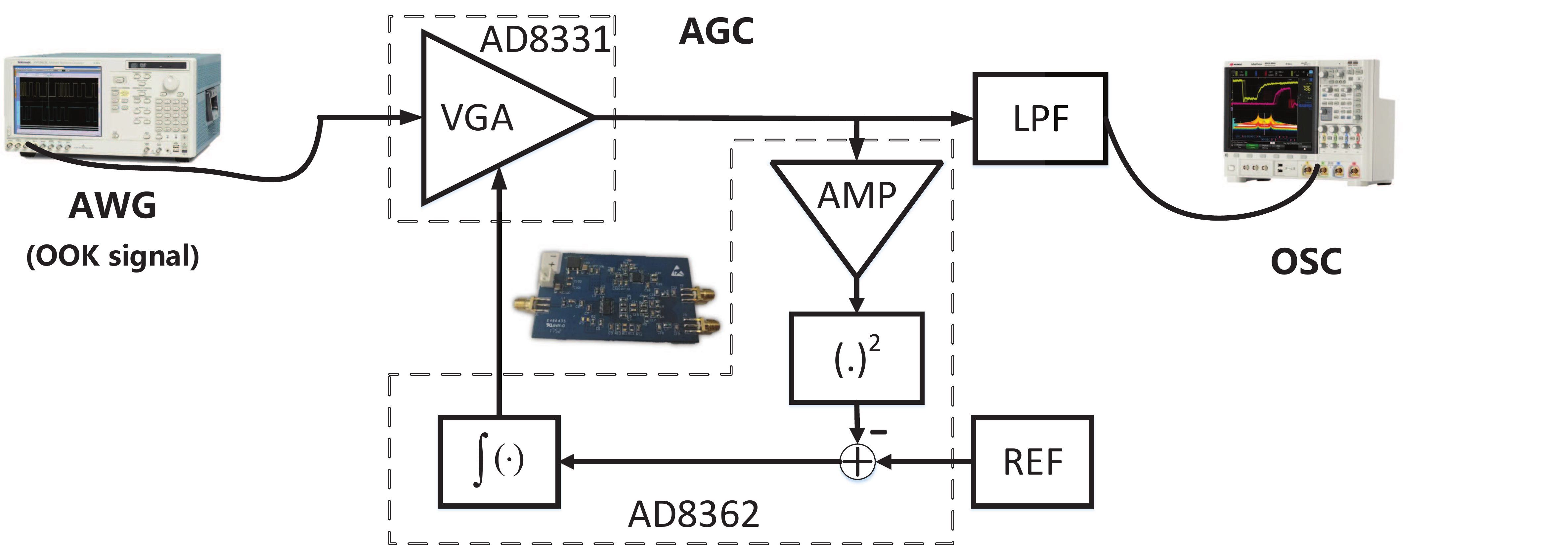}
	\end{minipage}}
	\subfigure[System diagram of real-time VLC mobile tracking platform]{
		\begin{minipage}[b]{0.9\columnwidth}
			\includegraphics[height=3.8cm]{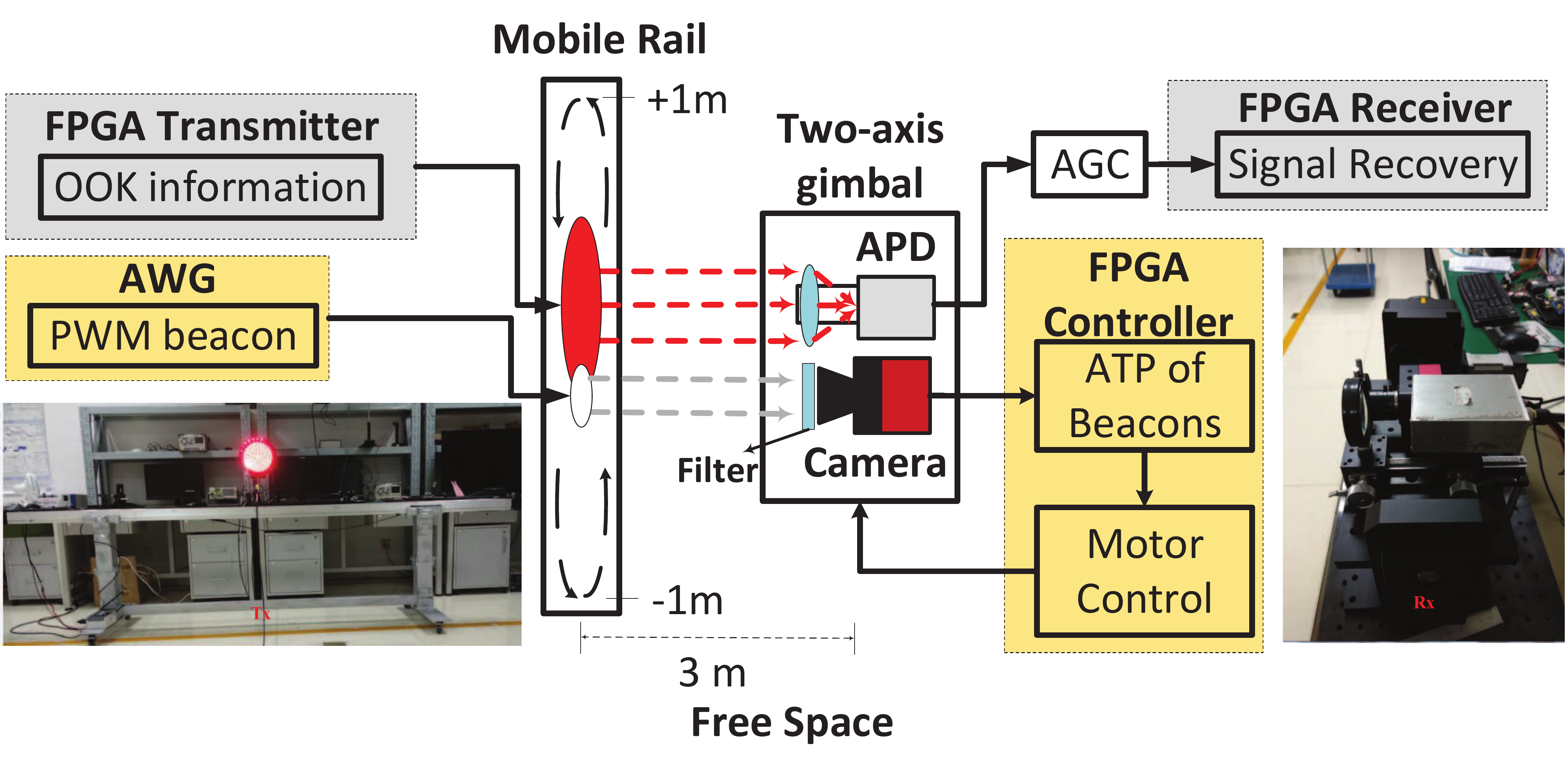}
	\end{minipage}} 
	\caption{Systems diagram for experiments } \label{fig:ExperimentSetup}
\end{figure}

\begin{figure}[!htbp]
	\begin{center}
		\includegraphics[height=6cm]{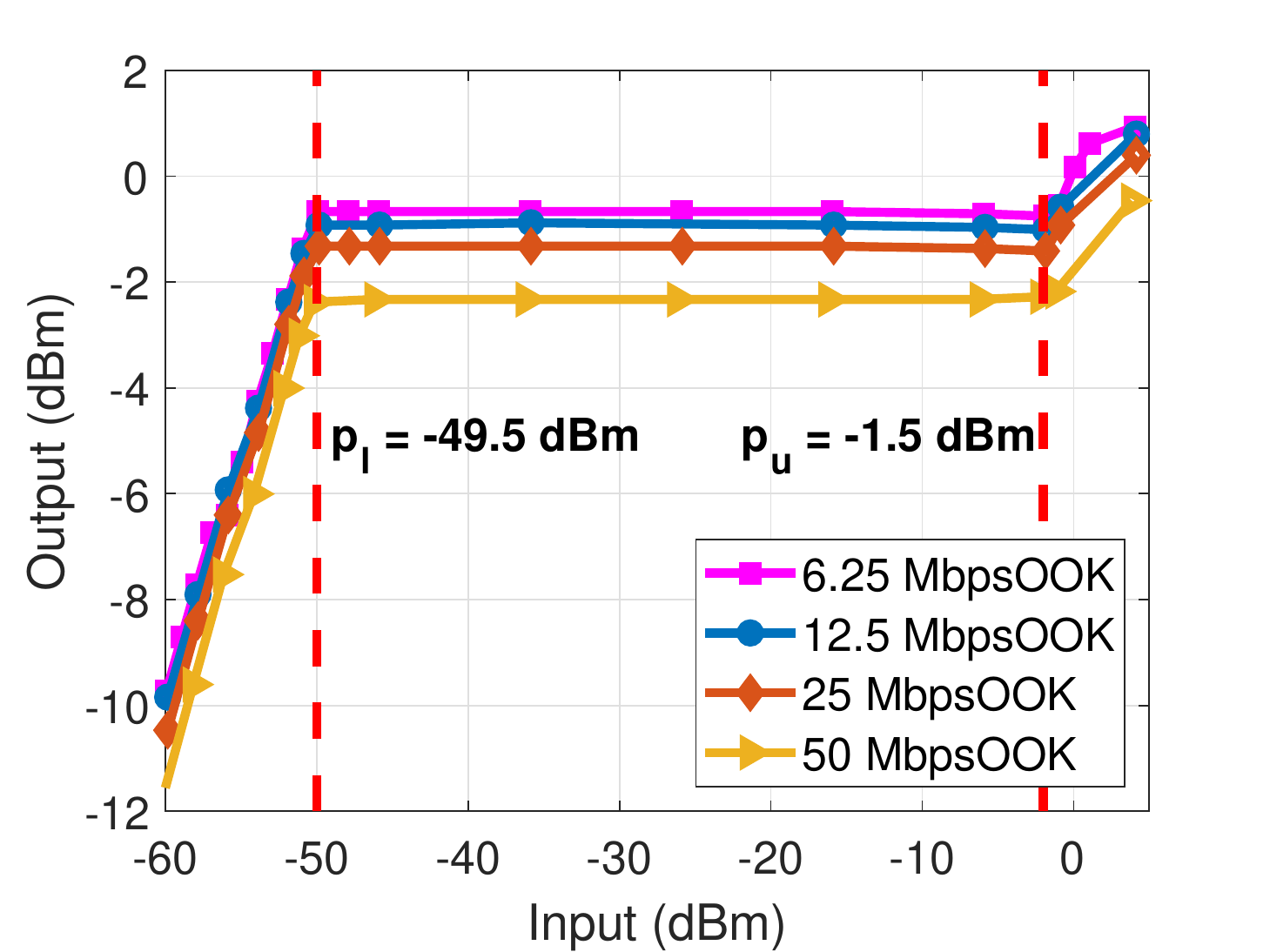}
		\makeatletter\def\@captype{figure}\makeatother
		\caption{Measured transmission characteristics for AGC amplifier}\label{fig:AGCtransCharac}
	\end{center}
\end{figure}

Numerous integrated analog AGC amplifiers are commercially available, and we implemented the circuit using AD8362 and AD8331 (from Analog Devices Inc.), to evaluate the effectiveness. The gain range for our VGA is -4.5 dB to 43.5 dB, and the AGC equilibrium output power was set as 0 dBm approximately. The AGC transmission characteristics with OOK signal was given in Fig. \ref{fig:AGCtransCharac}. The OOK signal was generated by arbitrary waveform generator (AWG), and we measured the AGC output directly using oscilloscope in 50 $\Omega$ (see experiment setup in Fig.\ref{fig:ExperimentSetup}.(a)). 

As depicted in Fig.\ref{fig:AGCtransCharac}, due to the gain of the extra loop amplifier (approximately 6dB), two threholds are $p_l=-49.5$ dB and $p_u=-1.5$ dB, respectively. On account of the frequency response of the chosen AGC components, the equilibrium power declines with the increasing bandwidth of input signals. While $p_{x}<p_1$, the AGC behaves as a fixed gain with maximum gain approximately 49.5 dB. On the other side, when $p_{x}>p_u$, we detected large harmonic distortions for this circuit, which makes the AGC unusable in this input region.   

\subsection{AGC experimental BER performance}
\begin{figure}[!htbp]	
	\begin{center}
		\includegraphics[height=6cm]{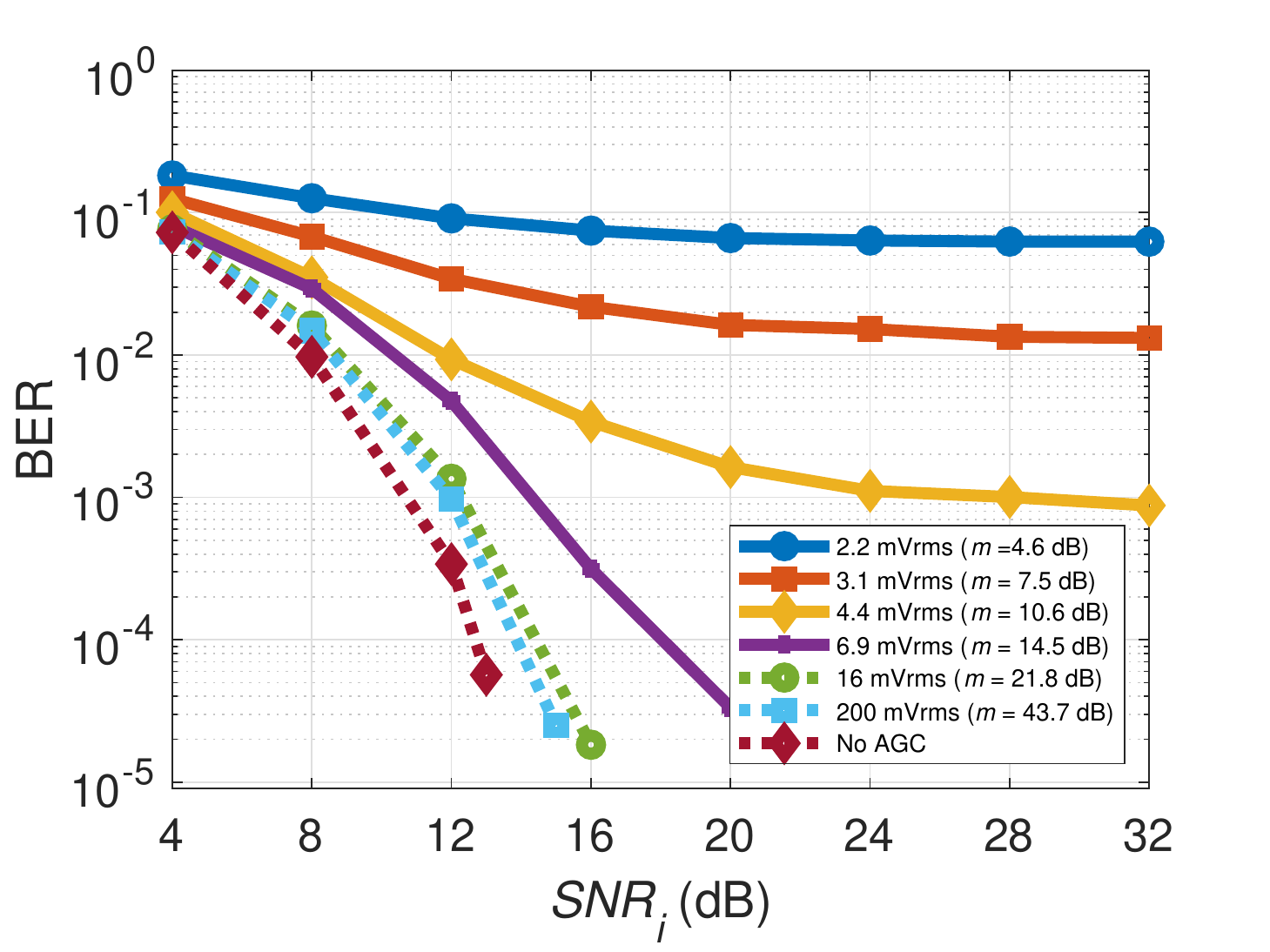}
		\makeatletter\def\@captype{figure}\makeatother
		\caption{Measured BER performance for different $p_e$}\label{fig:AGCberTest}
	\end{center}
\end{figure}

Next, to evaluate the AGC BER performance, a 1Mbps OOK signal, whose transmitted data is pseudo-random sequence, with different level Gaussian noise was intentionally generated by AWG. Through AGC amplifier, the output then was demodulated offline. While changing the AGC equilibrium power by adjusting the $v_{ref}$, the input signal amplitude was also adjusted to adapt to the AGC input thresholds. Using the AGC noise measured in Table \ref{tb:Simulation Parameters}, we further calculated $m$. In addition, the hard-decision demodulation (HDD) without AGC was conducted as the reference (results see Fig.\ref{fig:AGCberTest}).

In Fig.\ref{fig:AGCberTest}, coincided with our AGC model analysis earlier, in low SNR region ($SNR_i\ll m$), the output BER approaches to the HDD result (see curves of NoAGC, $m$=21.8 dB,43.7 dB), and while in high SNR region ($SNR_i\gg m$), certain BER floors occur (see curves of $m$=10.6 dB,7.5 dB,4.6 dB). Note that due to the AGC inherent non-linearity (non-linear gain, detector function and etc.), the signal quality is slightly deteriorated after passing AGC. This phenomenon can be effectively improved when applied equalization techniques.    
\subsection{Mobile tracking experiment with AGC}
Finally, the AGC function in mobile situation was tested in our real-time VLC mobile tracking platform\cite{liu2019simple} (system see Fig.\ref{fig:ExperimentSetup}.(b)). In this platform, the transmitting-end consists of a information LED, which used for data transmission, and a beacon LED, which used for position detection. The lights moved back and forth with different speeds on a rail. At the receiving-end, the light position was first detected by a high-speed camera using image processing technique and then sent to the motor to finish the light tracking process. Data demodulation was completed by APD. 

\begin{figure}[!htbp]
	\begin{center}
		\includegraphics[height=6cm]{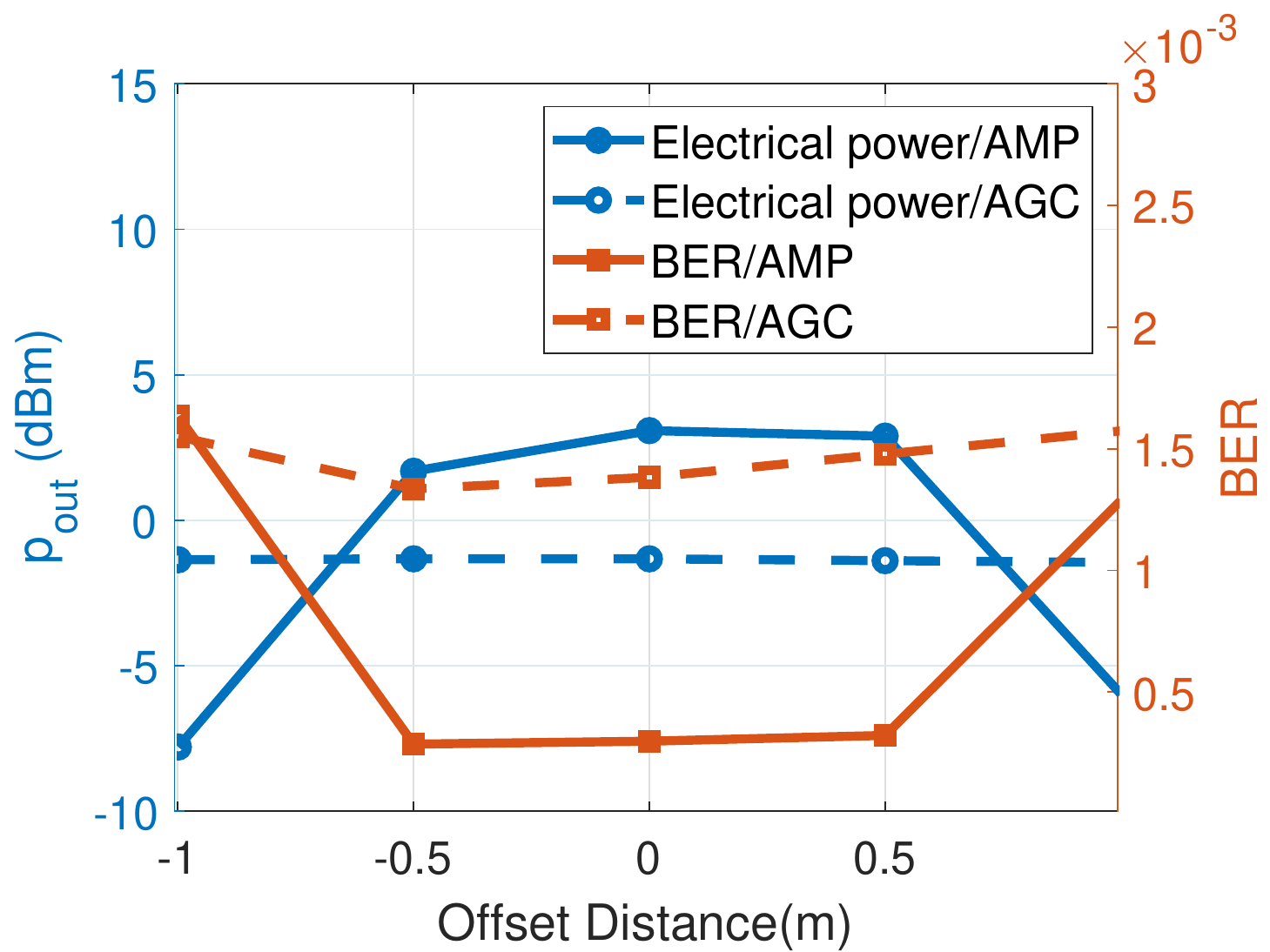}
		\makeatletter\def\@captype{figure}\makeatother
		\caption{Static BER benchmark}\label{fig:staticBenchmark}
	\end{center}
\end{figure}

We compared the AGC amplifier with a fixed gain amplifier with a power gain of 4.5 dB, where both the outputs of AGC and fixed gain amplifier are within the ADC input range. The static BER and output powers of two amplifiers at different rail position are given in Fig.\ref{fig:staticBenchmark} as benchmark. The transmitted signal is 25 Mbps OOK signal of 25 bits pseudo-random sequence and the communication distance is 3 m, with the AGC index $m$ = 43.7 dB.
\vspace{-0.3cm}
\vspace{-0.3cm}

As shown in Fig.\ref{fig:staticBenchmark}, when the transmitter places from rail sides to the rail center, the AGC is in equilibrium, and both the BER and output power stay relatively stable. The slight BER and power asymmetry phenomena arise from the asymmetry placements of the experimental equipment. 

\begin{figure}[!htbp]
	\begin{center}
		\includegraphics[height=6cm]{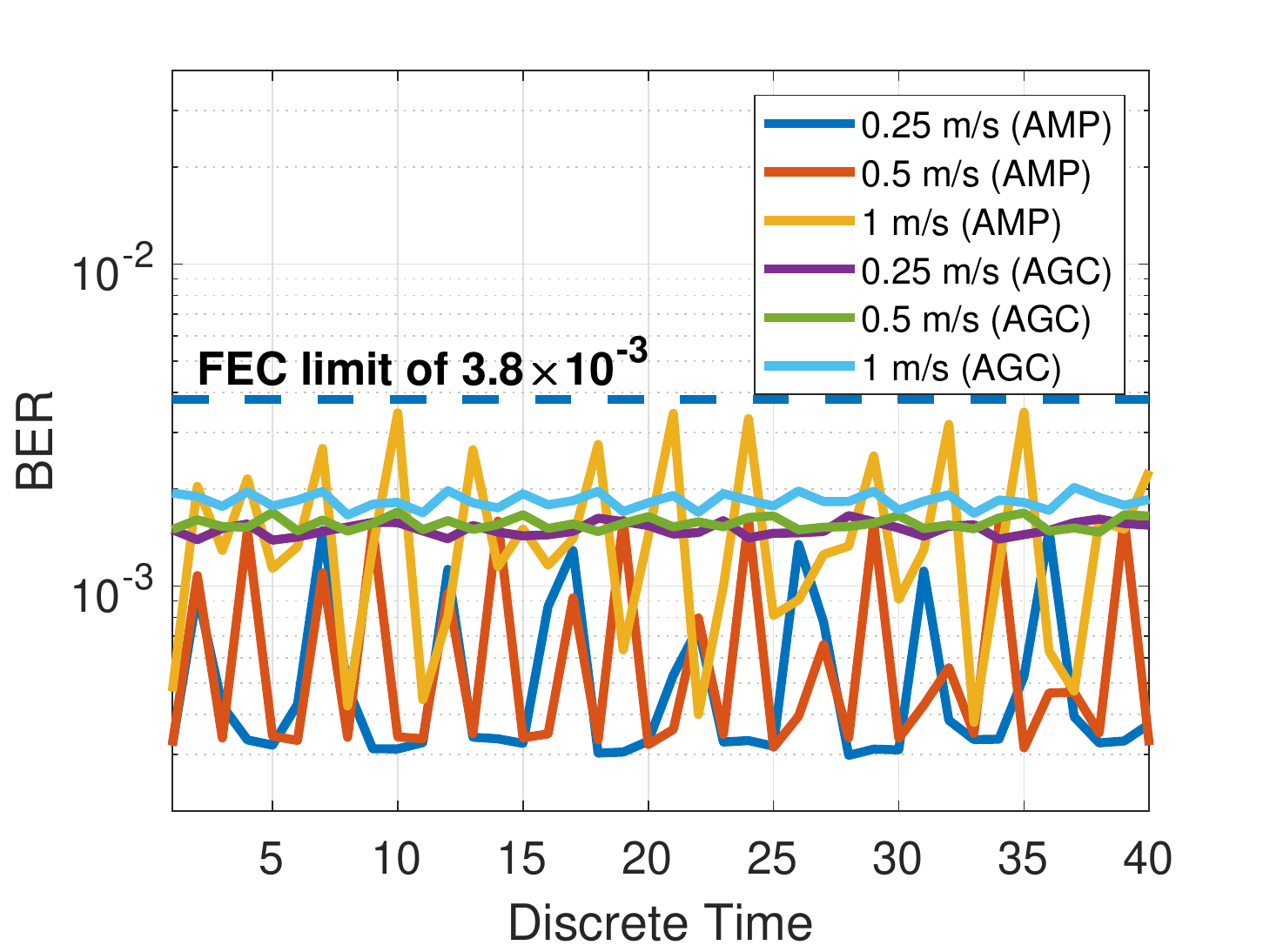}
		\makeatletter\def\@captype{figure}\makeatother
		\caption{Tracking BER performance at 25Mbps}\label{fig:moblieAGC}
	\end{center}
\end{figure}

Considering the values of $m$ and the AGC BER results, we know that the input SNR must be in low SNR region ($SNR_i\ll m$), but still the BER floor occurs. We may explain this as the effect of transmitted noise or LED non-linearity, since the non-linearity can be considered as one type of noise. Also, for the fixed gain amplifier, as discussed earlier, we ascribed the BER loss to the small gain. Real-time tracking experiments at moving speeds of 0.25 m/s, 0.5 m/s and 1 m/s were further performed to evaluate the AGC dynamic performance (see Fig.\ref{fig:moblieAGC}). 

As demonstrated in Fig.\ref{fig:moblieAGC}, when moving speeds are low (cases of 0.25 m/s and 0.5 m/s), the BER results are close to the static benchmarks. However, as the transmitter moving speed is up to 1 m/s (tracking angular speed of 18\textdegree/s), due to the mechanical delay of tracking, the light was slightly out of the plane of the APD, causing the power loss, thus BER degradation. 

Also note that, since the rise time of our AGC loop, provided by chip datasheet, is roughly 1ms (time of rise from 10\% to 90\% for a step input). And the power changing period in the 1 m/s case is 4 s. Thus the effect of loop dynamic setting is negligible. In practical vehicular scenario, one should also be considered is the power fluctuation caused by road irregularities and vehicle vibrations, whose vibration frequency are mainly low frequency component less than 20Hz\cite{kinoshita2014motion}.    
\vspace{-0.3cm}
\vspace{-0.6cm}
\section{Conclusion}
In this paper, we investigate the effect of AGC amplifier, both theoretical and experimentally.
Our key contributions and findings are as following: We modeled the VLC transmitter as a noisy transmitter, which can explain the BER floor in the experiment while the received signal power varied. A theoretical SNR model for the settled AGC was derived. On the basis of that, the effect of AGC index $m$ and the VGC gain $g_{max}$ were studied: 

Firstly, when the AGC is in the equilibrium-state, the output SNR is monotonically increasing with respect to $m$. In the high SNR region ($SNR_i\gg m$), a certain SNR floor occurs and in the low SNR region ($SNR_i\ll m$), the input SNR approaches the output SNR. We also pointed out a possible trade-off between the equilibrium output $p_e$ and the equilibrium thresholds $[p_l,p_u]$ while adjusting $m$.
Secondly, we found that proper VGA gain $g_{max}$ designing, namely, $g_{max}\gg p_{a}/\sigma_i^2r_l$, can avoid the undesired output SNR loss throughout the interval of $p_x<p_l$.  

Mobile VLC tracking experiments with AGC was conducted, which proved the effectiveness of AGC in dynamic scenario. Practical considerations such as the time constant of AGC was also mentioned, e.g., power fluctuations caused by road irregularities and vehicles vibrations.

{\small
\bibliographystyle{ieee_fullname}

}
\end{document}